\documentclass[
  ,final            
  ,numberedheadings 
  ]
  {aipproc}

\layoutstyle{6x9}

\newcommand{\hII}{H$^{\rm +}$}
\newcommand{\heII}{He$^{\rm +}$}
\newcommand{\heIII}{He$^{\rm ++}$}
\newcommand{\oII}{O$^{\rm +}$}

\begin{document}

\title{Primary neutral helium in the heliosphere}



\author{Hans-Reinhard~M\"uller}{
  address={Department of Physics and Astronomy, Dartmouth College,
Hanover NH 03755, USA}
 ,altaddress={Center for Space Plasma and Aeronomic Research,
  Univ.~of Alabama, Huntsville AL 35805, USA} 
}

\author{Jill~H.~Cohen}{
  address={Department of Physics and Astronomy, Dartmouth College,
Hanover NH 03755, USA}
}

\begin{abstract}
Two years of neutral measurements by IBEX-Lo have yielded several
direct observations of interstellar neutral helium and oxygen
during preferred viewing seasons. Besides the interstellar signal,
there are indications of the presence of secondary neutral helium
and oxygen created in the heliosphere. Detailed modeling of these
particle species is necessary to connect the measured fluxes to
the pristine local interstellar medium while accounting for loss
and production of neutral particles during their path through the
heliosphere. In this contribution, global heliosphere models are
coupled to analytic calculations of neutral trajectories to obtain
detailed estimates of the neutral distribution function of primary
interstellar helium atoms in the heliosphere, in particular in the
inner heliosphere.
\end{abstract}

\maketitle


\section{Introduction}

The neutral flux maps accumulated by NASA's Interstellar Boundary
Explorer (IBEX) since the start of its science operations in early
2009 contain detections of direct interstellar neutral helium and
oxygen atoms \cite{Moebius09}. These detections are centered
around time periods (in Spring) when IBEX scans in the direction
of the difference vector between Earth's orbital velocity and the
gravitationally deflected and accelerated particle velocity. The
main particle streams are consistent with a far-away velocity (in
the local interstellar medium, LISM) determined by \cite{Witte04}
who obtained the helium flow field by careful analysis of the
Ulysses/GAS measurements (see also \cite{Moebius04} for a synopsis
of the values from different techniques). Further analysis of the
IBEX data reveals the presence of secondary oxygen and secondary
helium in the measurements. Those particles are created through
charge exchange of \heII, \heIII, or \oII\ with heliospheric
neutrals (H, He, O) and hence are in a different velocity state,
with different neutral trajectories.

Calculating primary and secondary neutral fluxes at IBEX through
modeling requires combining a global heliosphere model treating
the dominant neutral H and \hII\ (protons) self-consistently to
obtain a realistic heliospheric plasma background. On this
background, heavy neutrals like He and O can be simulated with
Monte-Carlo-style kinetic codes \citep[e.g., ][]{Mueller04} where
primary interstellar neutrals are injected at an outer boundary,
sampled from a thermal Maxwellian, and transported through the
heliosphere while subjected to loss processes (interaction with
the background plasma and solar photons). Secondary neutrals are
generated by charge exchange everywhere in the heliosphere, and
after creation they are transported as well, while subject to
losses.

This paper pursues an alternative method for calculating heavy
neutrals, using Keplerian equations to describe the entire
trajectory while conserving quantities like energy and angular
momentum to maximize the accuracy of the results. Using conserved
quantities connects a 1 AU location (IBEX) with the pristine ISM
(LISM) in one algebraic step. With this efficient method, the goal
then is to carry out precision calculations, where the entire
phase space distribution function at a desired point is calculated
as accurately as desired without the steep computational cost that
a Monte Carlo method typically incurs for a similar accuracy.

\section{Conserved quantities along trajectory}

The Sun exerts a central force on neutral particles in the
heliosphere, and the neutral trajectories hence are Keplerian
orbits with the Sun at the focus. The motion proceeds in a central
potential -$G M_{\bigodot}/r$. More generally, the potential
constant is defined as $f_{\mu} = G M_{\bigodot}(1-\mu)$, with $G$
the gravity constant, $M_{\bigodot}$ the solar mass ($G
M_{\bigodot} = 887.6$ AU km$^2$ s$^{-2}$ is used here), and $\mu$
the ratio of radiation pressure force to gravitational force.
While $\mu\approx 1$ for hydrogen, $\mu =0$ for heavy neutrals,
with solar gravity acting unopposed.

The motion of an individual neutral particle is confined to a
plane (orbital plane), containing the Sun, the particle, and its
velocity vector. There are several conserved quantities along the
entire trajectory. The most important ones are the following
seven, which are not all independent from each other: Angular
momentum, Laplace-Runge-Lenz vector, and total energy. As those
all involve the particle mass or its square as prefactors, the
latter can be made part of the conservation constants, to arrive
at the following set of conserved quantities:

\begin{eqnarray}
\vec l & = & \vec r \times \vec v \mbox{\ \ \
    (specific angular momentum)}\label{eqangmom}\\
\vec a & = & \vec v \times \vec l - f_{\mu} \hat{r} \mbox{\ \ \
   (eccentricity vector)}\label{eqecc}\\
E_{tot} & = & {v^2\over 2} - {f_{\mu} \over r} \mbox{\ \ \
   (total specific energy)} \label{eqetot}
\end{eqnarray}
With an appropriate coordinate system, two angular momentum
components and one eccentricity vector component can be made to
identically vanish everywhere. The eccentricity vector $\vec a$
points in the direction of perihelion, and its magnitude is
related to the orbital eccentricity $e$ by $|\vec a| = e \,
f_{\mu}$.

The conserved quantities can be used in at least two ways: In a
direct way, when a particle's location and velocity is specified
at one time. Then, all conserved quantities are immediately
determined, and the entire future (forward) trajectory and past
(reverse) trajectory can be calculated with formulae from
celestial mechanics. With the knowledge of the conserved
quantities, any desired radial distance, azimuthal angle (``true
anomaly''), point in time, or radial velocity, for example, can be
used as input to determine all other variables accurately, with
the conserved quantities conserved by design. A natural
application of this is backtracking a particle to a large distance
where it is sure to be in the LISM. In this way, a particle
measured by IBEX with a certain direction and energy can be
backtracked to where it came from, and the travel time from a
reference distance to IBEX can be calculated algebraically as
well.

In the second way, the mixed problem of knowing a particle's
position at one point in time, and a particle's velocity vector at
infinite distance where it started out, is solvable as well. Here,
two sets of conservation equations (\ref{eqangmom})-(\ref{eqetot})
are established, and treated as a system of equations that, when
solved, yields the missing variables. An application of this
problem has been published quite some time ago in the context of
characterizing the velocity state of neutrals in the inner
heliosphere when a cold interstellar neutral distribution (known
velocity at infinity) is assumed \cite{Axford72}. Equation 48 of
\cite{Axford72} gives the algebraic formula for the angular
momentum and hence velocity vector at the location of interest.
Each location is reached by two neutral trajectories (two
solutions), a direct path and a longer indirect path. Typically,
the indirect path takes the particle closer to the Sun, and for a
longer time, and is hence more affected by losses than the direct
path; see Figure 2a below for an illustration which also contains
the respective values for eccentricity and specific angular
momentum (the latter in units of AU km/s).

\section{Gravitationally deflected phase space density}

In the pristine LISM a Maxwellian phase space distribution (PSD)
can be assumed for the interstellar neutral helium atoms. Through
the action of gravity, the PSD shape at a point in the inner
heliosphere is different from Maxwellian, even if charge-exchange
collisions are ignored. The PSD maximum is determined by using the
bulk speed vector of the LISM \cite{Witte04} as input and solving
the mixed problem as outlined above. The entire PSD is obtained by
scanning the velocity space around this maximum in all velocity
space directions; for each new velocity, this ``direct problem''
(see above) is solved, backtracking the particle to yield the
velocity at infinity. By virtue of Liouville's theorem, the PSD of
the Maxwellian for this velocity equals the one of the phase space
point under investigation. For practical purposes, such scanning
is ended when the PSD value falls below $10^{-4}$ of the maximum
value, which is on the order of 4 thermal velocities away from the
bulk speed in the LISM.

\begin{figure}
\centering
\includegraphics[width=0.95\textwidth]{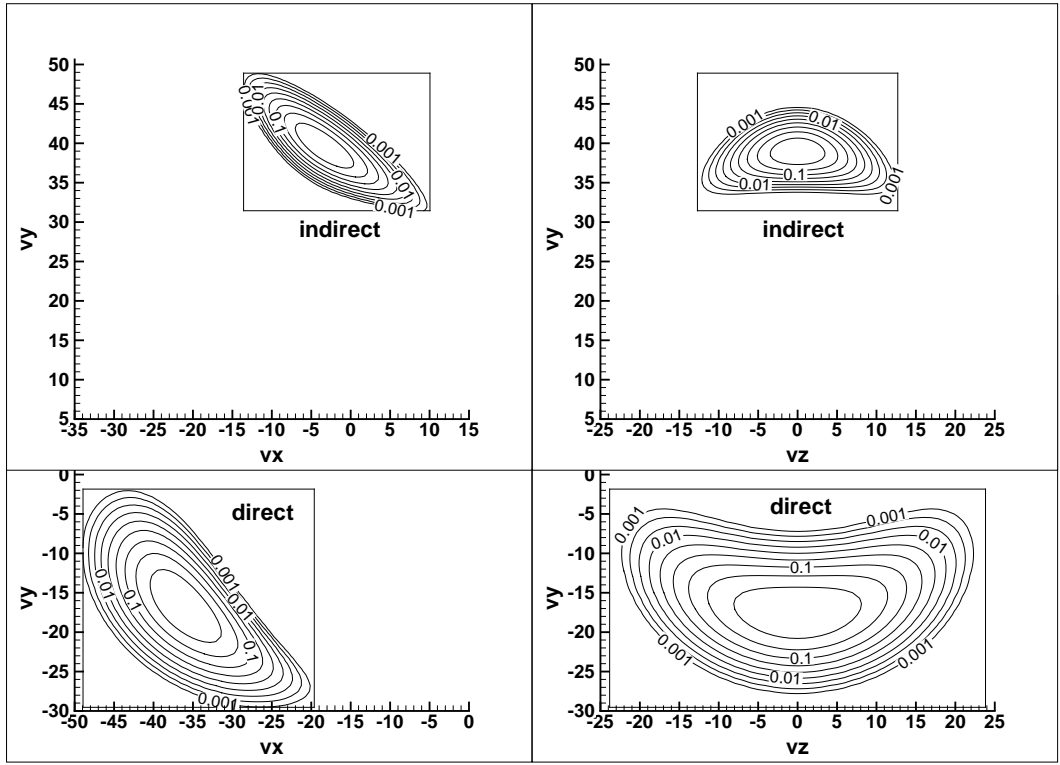}
\caption{Phase space densities of primary neutral interstellar
helium at the location (-1.0, 1.7) slightly downwind of the Sun
and away from the symmetry axis (2 AU distance). The PSD is a 3D
object; the left column presents $v_x$-$v_y$ cuts through the
center ($v_z$=0 in both panels; note the $v_x$ axis shift between
the panels). Analogously, $v_y$-$v_z$ cuts taken at the centers
$v_x$=-3 km/s (indirect PSD) and at $v_x$=-36 km/s for the direct
PSD constitute the right column of the figure. The maximum value
is normalized to 1 (no losses included); the contour levels are
spaced logarithmically with 3 lines per dex.\label{fig1}}
\end{figure}

Consider an example of this procedure, with a sample point of
interest chosen 2 AU away from the Sun, in a downwind-sidewind
direction that is not yet in the helium focusing cone. The
location coordinates $\vec r$=(-1.0, 1.73, 0.0) in a frame of
reference where the $x$ axis points into the LISM flow, combined
with the LISM bulk velocity (-26.3, 0.0, 0.0) km/s \cite{Witte04},
give the direct-path velocity solution (-36.0, -16.8, 0.0) km/s
and the indirect solution (-3.4, +39.6, 0.0) km/s. The situation
is illustrated in Figure 2a. In the entire velocity space at $\vec
r$, there are two maxima of value 1 at these two velocities, each
surrounded by a finite volume where the normalized phase space
density remains larger than 0.001, an arbitrary value chosen to
envelop the most important pieces of phase space. Figure 1 shows
perpendicular cuts through these two 3D velocity space objects,
all taken at planes that contain the respective point of maximum.
The shape of these cuts suggests that the deflected direct PSD has
a form similar to a thick contact lens; the indirect PSD is more
irregular yet confined to a smaller volume. For comparison, the
LISM Maxwellian contours are concentric circles, with the 0.001
contour having a radius of 14 km/s.

The procedure so far disregards any losses interstellar helium
might experience on its way from the pristine LISM to a point in
the inner heliosphere. Photoionization is the dominant loss; its
local rate is proportional to the inverse square of distance to
the Sun. Charge exchange (c.x.) channels are the next most
important losses; their rate is proportional to the background
plasma density, the relative plasma-neutral speed, and c.x. cross
section. The c.x. channels are loss of helium due to encounters
with plasma protons, with \heII\ ions in the outer heliosheath,
and with \heIII\ ions (alpha particles) in the solar wind. Some
reasonable abundances \cite{Slavin08} are chosen to represent
helium ion densities and velocities with the help of the proton
plasma background of an axisymmetric global heliosphere model. The
introduction of the numerical plasma background presents an
increase in the amount of calculations: Now, the orbital position
and velocity at each background grid point need to be calculated
to evaluate the local loss rate. With the conserved quantities
available, this still is a one-step calculation; it is just the
number of such calculations that goes up, roughly to the order of
one linear dimension of the background plasma grid.

\begin{figure}
\centering
\includegraphics[width=0.373\textwidth]{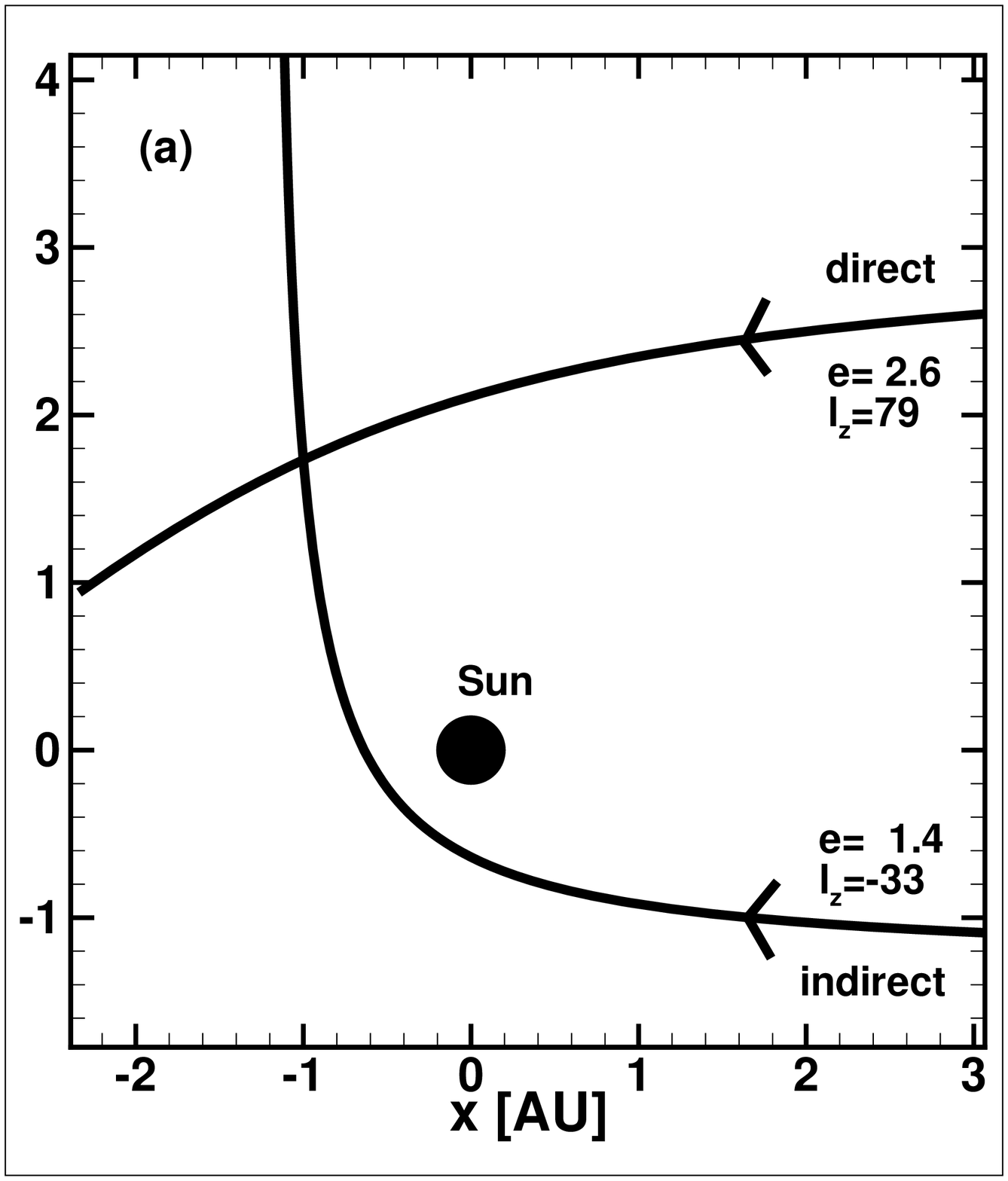}
\includegraphics[width=0.60\textwidth]{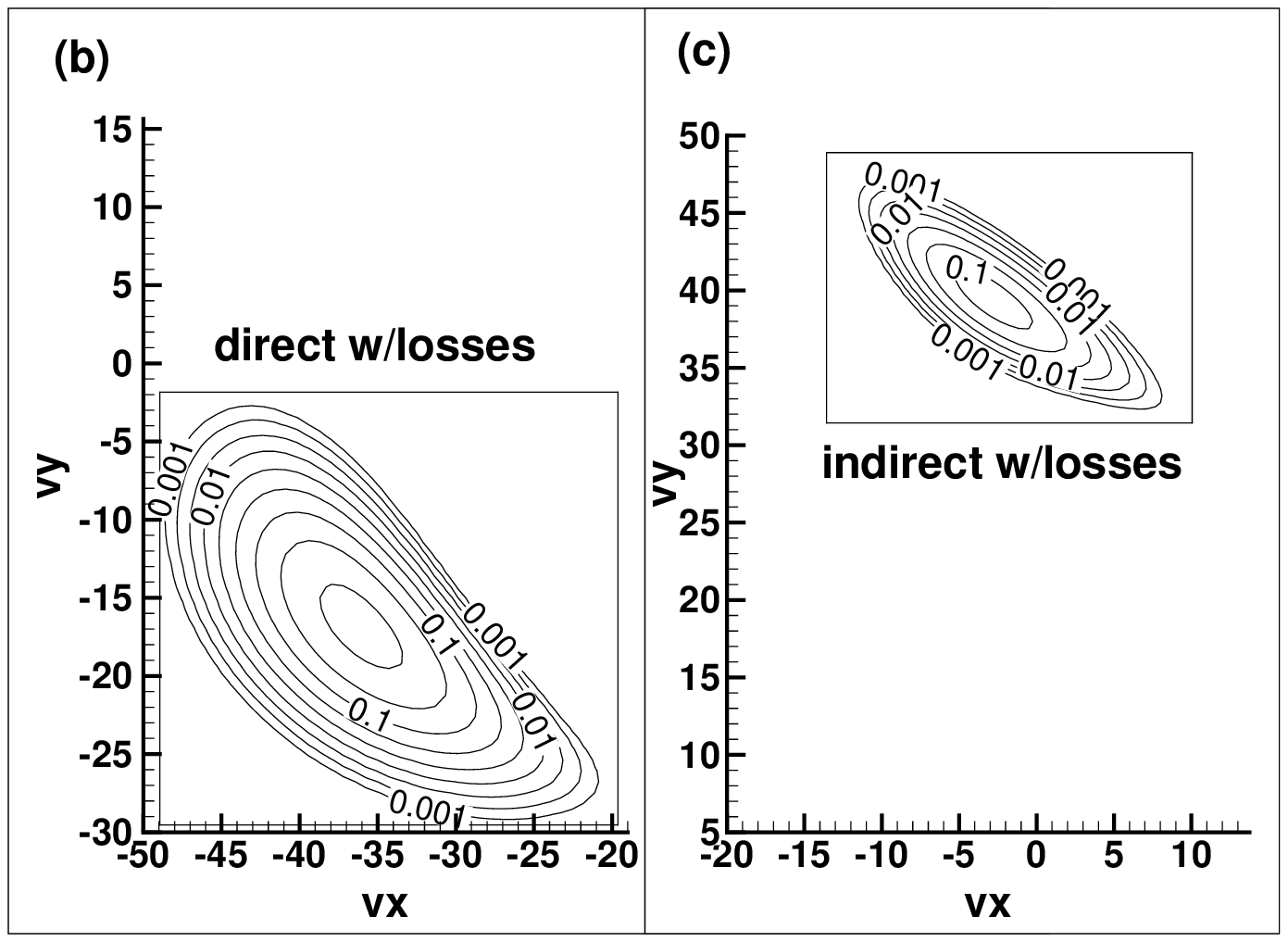}
\caption{(a) Schematic of direct and indirect paths reaching the
point (-1, 1.7). (b, c) $v_x$-$v_y$ cuts through the phase space
density after losses have been taken into account. See text for
details. The contour levels are identical to those in Figure
1.\label{fig2}}
\end{figure}

In keeping with the above example, Figure 2 shows the modified
$v_x$-$v_y$ cuts of the PSD when all four loss processes are taken
into account. The direct PSD is lower now, with a new maximum of
0.61 (the first contoured level is 0.46). The PSD has a smaller
slope now, so that the 0.001 contour location is not very
different from before. In contrast to this, the indirect path
takes the particle closer to the Sun, and the corresponding PSD
region suffers more dramatic losses, with even the 0.001 contour
surrounding a much smaller volume. The maximum achieved at the
center is 0.14, indicative of strong loss.

With the PSD characterized in such a detailed fashion, it is easy
to integrate moments to arrive at primary neutral density, bulk
velocity, and temperature, also separate for direct and indirect
populations if desired. The PSD in the wake of the Sun, where a
focusing cone exists, has a complex shape, degenerating into a
ring on the downwind symmetry axis. There, higher moments do not
characterize the actual neutral PSD well.

To complete the picture and make detailed connections with
observations, it will be necessary to add secondary neutral
particles to the calculations. They are the neutral product of
charge exchange of a helium ion with a neutral H atom (supplied by
a background global heliosphere model) or a neutral He atom (as
calculated above). He ions are part of the solar outflow and the
ISM inflow, as mentioned above for the c.x. loss processes of
primaries. Naturally, the pickup ions generated in primary c.x.
events are a further source of He ions available for the
production of secondary neutral He. It is often assumed that the
level of pickup ions is negligible when contrasted with the solar
wind and interstellar helium ion densities, but this should be
verified in detail for typical global heliospheric models. In case
the pickup ions contribute significantly to the secondary helium,
techniques for their inclusion in the global models will include
location-dependent extra source terms for the helium ion fluid.

While the source of primary helium is at infinity, the source of
secondary helium leading to a particular phase space point is
distributed along the entire trajectory. Therefore, the secondary
PSD will involve by necessity an integral along the trajectory,
with primary He calculations at each point of the trajectory. In
this sense, non-primary particle calculations are a hierarchical
process where ``higher orders'' are derived from ``lower orders'',
and therefore call for more involved calculations. These issues
will be treated in future papers.

\section{Conclusions}

Using the Kepler equations of celestial mechanics is advantageous
when calculating trajectories in the heliosphere of neutral heavy
atoms originating in the local interstellar medium. The associated
conserved quantities enable efficient, one-step calculations of
particle locations and velocities that are accurate by design.

One example was presented to show the accurate output of the
method, emphasize the role of gravitational deflection on the
helium phase space distribution, and discuss the changes that
further result from loss processes. There are two maxima in the
helium PSD in the heliosphere, but typically the indirect path
contribution suffers more severe losses.

The primary neutral helium PSD can be calculated in this way at
any arbitrary point in the heliosphere, and its various moments
will yield maps of effective temperature, bulk velocity, and
density. The latter will be elevated in the region of the helium
focusing cone, which is also where direct and indirect solutions
start to merge together and therefore become equally important.
The general helium filtration by the heliospheric interface, by
solar radiation and the inner solar wind is more pronounced than
earlier studies suggested (e.g., \cite{Mueller04}) in which only
charge exchange with protons was taken into account as loss
process. These issues will be studied in more detail elsewhere.


\begin{theacknowledgments}
Partial support of this work by NASA grants NNX10AC44G,
NNX11AB48G, NNX10AE46G, and by a University of Chicago subcontract
of NASA grant NNG05EC85C is gratefully acknowledged. HRM thanks
Vladimir Florinski, Priscilla Frisch, and Gary Zank for helpful
interactions.
\end{theacknowledgments}



\bibliographystyle{aipproc}   

\begin{thebibliography}{6}
\expandafter\ifx\csname
natexlab\endcsname\relax\def\natexlab#1{#1}\fi
\providecommand{\enquote}[1]{``#1''} \expandafter\ifx\csname
url\endcsname\relax
  \def\url#1{\texttt{#1}}\fi
\expandafter\ifx\csname
urlprefix\endcsname\relax\def\urlprefix{URL }\fi
\providecommand{\eprint}[2][]{\url{#2}}

\bibitem[{M{\"o}bius} et~al.(2009)]{Moebius09}
E.~{M{\"o}bius}, P.~{Bochsler}, M.~{Bzowski}, G.~B. {Crew}, H.~O.
{Funsten},
  S.~A. {Fuselier}, A.~{Ghielmetti}, D.~{Heirtzler}, V.~V. {Izmodenov},
  M.~{Kubiak}, H.~{Kucharek}, M.~A. {Lee}, T.~{Leonard}, D.~J. {McComas},
  L.~{Petersen}, L.~{Saul}, J.~A. {Scheer}, N.~{Schwadron}, M.~{Witte}, and
  P.~{Wurz}, \emph{Science} \textbf{326}, 969--971 (2009).

\bibitem[Witte(2004)]{Witte04}
M.~Witte, \emph{Astron. Astrophys.} \textbf{426}, 835--844 (2004).

\bibitem[{M{\"o}bius} et~al.(2004)]{Moebius04}
E.~{M{\"o}bius}, M.~{Bzowski}, S.~{Chalov}, H.-J. {Fahr},
G.~{Gloeckler},
  V.~{Izmodenov}, R.~{Kallenbach}, R.~{Lallement}, D.~{McMullin}, H.~{Noda},
  M.~{Oka}, A.~{Pauluhn}, J.~{Raymond}, D.~{Ruci{\'n}ski}, R.~{Skoug},
  T.~{Terasawa}, W.~{Thompson}, J.~{Vallerga}, R.~{von Steiger}, and
  M.~{Witte}, \emph{Astron. Astrophys.} \textbf{426}, 897--907 (2004).

\bibitem[M{\"u}ller and Zank(2004)]{Mueller04}
H.-R. M{\"u}ller, and G.~P. Zank, \emph{J. Geophys. Res.}
\textbf{109}, A07104
  (2004).

\bibitem[Axford(1972)]{Axford72}
W.~I. Axford, \enquote{The interaction of the solar wind with
interstellar
  medium,} in \emph{Solar Wind}, C.~P. Sonett, P.~J. Coleman, and
  J.~M. Wilcox (Eds.), NASA SP-308, 609--660 (1972).

\bibitem[{Slavin} and {Frisch}(2008)]{Slavin08}
J.~D. {Slavin}, and P.~C. {Frisch}, \emph{Astron. Astrophys.}
\textbf{491}, 53--68 (2008).

\end{thebibliography}

\hyphenation{Post-Script Sprin-ger}

\end{document}